\documentclass[preprint,showpacs,preprintnumbers,amsmath,amssymb,nofootinbib]{revtex4}

\usepackage{graphicx}
\usepackage{dcolumn}
\usepackage{bm}

\newcommand{\bi}{\bibitem}
\newcommand{\nn}{\nonumber}

\newcommand{\be}{\begin{eqnarray}}
\newcommand{\ee}{\end{eqnarray}}
\catcode`\@=11
\def\lsim{\mathrel{\mathpalette\@versim<}}
\def\gsim{\mathrel{\mathpalette\@versim>}}
\def\@versim#1#2{\vcenter{\offinterlineskip
\ialign{$\m@th#1\hfil##\hfil$\crcr#2\crcr\sim\crcr } }}
\catcode`\@=12

\begin{document}

\preprint{KANAZAWA-05-01}

\title{$S_3$ Flavor Symmetry and Leptogenesis}

\author{Takeshi Araki$^{*}$}
\author{Jisuke Kubo$^{*}$}
\author{Emmanuel A. Paschos$^{**}$}
\affiliation{$^{*}$
Institute for Theoretical Physics, Kanazawa
University, Kanazawa 920-1192, Japan\\
$^{**}$Institut f\"ur Physik, Universit\"at Dortmund,
Otto-Hahn-Str.4, 44221 Dortmund, Germany}

\begin{abstract}
We consider leptogenesis in a minimal
 $S_3$ extension of the standard model
with an additional $Z_2$ symmetry in the leptonic sector.
It is found that  the CP phase appearing
in the neutrino mixing is the same
as that for the CP asymmetries responsible for leptogenesis.
Because of the discrete $S_3\times Z_2$ flavor symmetries,
the CP asymmetries 
are strongly suppressed.
We therefore assume that the resonant enhancement of the 
CP asymmetries takes place
to obtain a realistic size of baryon number asymmetry in the
universe.
Three degenerate right-handed neutrino masses of $O(10)$ TeV
are theoretically expected in this model.
 
\end{abstract}


\maketitle

\section{Introduction}
 The baryon asymmetry in the universe 
 brings cosmology and particle physics together \cite{kolb}.
A theoretically attractive idea  \cite{fukugita1} to produce
 baryon asymmetry is to apply 
 a nonperturbative conversion mechanism of 
lepton asymmetry to  baryon asymmetry, which exists
as the sphaleron process  in  the standard model (SM) \cite{manton,kuzmin}.
For this idea to work, a sufficient amount of
$B-L$  has to be generated \cite{khlebnikov}
at temperatures $T$ between  $100$  and $10^{12}$ GeV \cite{arnold,bochkarev}.
With the experimental fact that the neutrinos are 
massive \cite{superK,kamL,k2k,sno},
it is plausible to believe that they are Majorana particles and hence
the lepton number conservation is violated.
This situation is  nicely  realized  in the see-saw mechanism \cite{seesaw},
which, after spontaneous symmetry breaking of $SU(2)_L\times U(1)_Y$,
generates the Majorana masses of the left-handed neutrinos
in the presence of heavy right-handed neutrinos.

With the see-saw mechanism at hand, it becomes indeed possible to
explain the observed ratio of baryons to photons
$\eta_B = (6.2-6.9) \times 10^{-10}$ \cite{wmap} 
by leptogenesis 
\cite{luty}-\cite{pilaftsis2}. However, the introduction of 
the right-handed neutrinos into the SM introduces
additional ambiguities in the Yukawa sector.
Because of these ambiguities, 
the theoretical value of  $\eta_B$
depends on many independent parameters,
so that it would be very difficult to make quantitative tests
of the different mechanisms involved to produce baryon asymmetry.
The origin of these ambiguities in the Yukawa sector
in the SM is the  missing of a more strict theoretical guide
to construct the Yukawa sector.

A natural guidance to constrain the Yukawa sector
is a flavor symmetry.
Although there are attractive continues
symmetries, we would like to consider
discrete symmetries, especially nonabelian discrete 
symmetries \footnote{ Earlier  papers
on permutation symmetries  are
\cite{pakvasa1,harari,derman,wyler,segre,sato,yamanaka,koide1} for instance.
See \cite{fritzsch} for a review.}.
However, experimental data require that within the framework of the SM
any nonabelian flavor symmetry has to be hardly  broken
at low energy.
If the Higgs sector of the SM is so extended that
a certain set of  Higgs fields belong to
a nontrivial representation of a nonabelian flavor group,
phenomenologically viable possibilities may arise \footnote{
Recent, phenomenologically
viable models based on 
nonabelian discrete flavor symmetries $S_{3}, D_{4}, A_{4}, Q_4$ and $Q_6$
and also on a product of abelian discrete symmetries 
 have been recently constructed in
 \cite{ma,kubo1,kubo2,kobayashi1,choi,ma5},  \cite{grimus2,grimus3},  
 \cite{ma1,babu1,babu2,hirsch},\cite{frigerio},\cite{babu4}
and \cite{grimus1,ma4,grimus6}, respectively.
 See also \cite{koide3}-\cite{low2}.}.
The  smallest nonabelian
discrete group is  $S_{3}$.
In this paper, we would like to consider a
minimal  $S_3$ invariant extension 
of the SM \cite{kubo1,kubo2},
in which $S_3$ is not hardly broken at
low energy, and consequently, the Yukawa sector is much more constrained than
in the SM. 
In Sect. II we define the model, while investigating the
independent phases in the leptonic sector.
We discuss neutrino mixing and CP phase in Sect. III,
and express the average
neutrino mass $<m_{ee}>$ appearing in neutrinoless double $\beta$ decay
as a function of the independent phase $\phi_\nu$.
Leptogenesis and baryon asymmetry are  considered in Sect. IV, while
Sect. V is devoted to summarizing our findings.

\section{$S_{3}$ invariant extension of the  standard model }
\subsection{Leptonic sector}
We assume that the three generations of quarks and leptons 
 belong to the reducible representation
of $S_3$ ${\bf 3}={\bf 1}+{\bf 2}$.
We also introduce an $S_3$ doublet Higgs fields,
$H_{i}(i=1, 2)$, as well as an
$S_3$ singlet Higgs, $H_{S}$.
The $S_{3}$ invariant Yukawa interactions in the leptonic sector 
is given by \cite{kubo1} \footnote{
In \cite{kubo1} it is incorrectly stated that
there is no CP phase in the present model.}
\be
{\cal L}_{Y} &=&
-y_{1}\overline{ L}_i H_S E_{iR} - y_{3} \overline{ L}_S H_{S} E_{SR}
-y_{2}f_{ijk}\overline{ L}_{i} H_j  E_{kR}\nn\\
& & - y_{4} \overline{ L}_S H_{i} E_{iR}
- y_{5} \overline{ L}_i H_{i}E_{SR}\nn\\
& &-h_{1} \overline{ L}_I \epsilon H_S^{*} \nu_{iR} 
-h_{3} \overline{ L}_S\epsilon  H_{S}^{*} \nu_{SR}
-h_{2}f_{ijk}\overline{ L}_{i} \epsilon H_{j}^{*}  \nu_{kR}\nn\\
& & - h_{4} \overline{L}_S \epsilon H_{i}^{*} \nu_{iR}
- h_{5} \overline{L}_i \epsilon H_{i}^{*} \nu_{SR}\nn\\
& &- \frac{1}{2}M_{1}\overline{ \nu}_{iR}\nu_{iR}-
 \frac{1}{2}M_{S}\overline{\nu}_{SR}\nu_{SR}+h.c.,~i,j,k=1,2,
\label{Ly}
\ee
where
\be
f_{121} &=& f_{211}=f_{112}=-f_{222}=1~,~
f_{111}=f_{221}=f_{122}=f_{212}=0.
\label{fijk}
\ee
Here $L,E_R,\nu_R$ and $H$ stand for  left-handed
charged lepton $SU(2)_L$ doublets, right-handed 
charged leptons, 
right-handed  neutrinos and Higgs  $SU(2)_L$ doublets,
respectively.
${\cal L}_{Y}$ is the most general renormalizable  form
that is $S_{3}$ invariant.
In \cite{kubo1} an additional abelian discrete symmetry $Z_2$
has been  introduced to achieve a further simplification
of the leptonic sector. The $Z_2$ parity assignment is:
\be
+&  & \mbox{ for} ~~H_i, ~L_S, ~L_i, ~E_{iR},~ E_{SR},~
\nu_{iR}
~\mbox{ and}~~
- ~~\mbox{for}~~H_S,~\nu_{SR}.
\label{z2}
\ee
This $Z_{2}$ 
forces the following Yukawa couplings
to vanish \footnote{This symmetry is hardly broken
 in the quark sector.
Therefore, there will be radiative corrections coming from
that sector. However, they appear first at the two-loop level, and
so, one may assume that they are small.}:
\be
y_{1},y_{3},h_{1}~\mbox{and}~h_{5}.
\ee

Let us next figure out the structure of CP phases.
To this end, we
introduce phases explicitly as follows:
\be
y_{a}&\to &e^{i p_{y_{a}}} y_{a} ~(a=2,4,5)~,~
h_{a} \to e^{i p_{h_{a}}} h_{a} ~(a=2,4,3)
\label{phases}
\ee
for the Yukawa couplings, where $y$'s and $h$'s 
on the right-hand side are assumed to be real and
$-\pi/2 \leq p's\leq \pi/2$, and similarly for the fields
\be
L_{i}&\to & e^{i p_{L}}L_{i}~,~
L_{S} \to e^{i p_{L_{S}}}L_{S}~,~
E_{iR}\to  e^{i p_{E}}E_{iR}~,~
E_{SR} \to e^{i p_{E_{S}}}E_{SR},\nn\\
\nu_{iR} &\to &  e^{i p_{\nu}}\nu_{iR}~,~
\nu_{SR} \to e^{i p_{\nu_{S}}}\nu_{SR}.
\ee
The phases of the right-handed neutrinos 
are used to absorb the phase of their Majorana
masses $M_{1}$ and $M_{S}$.
The phases of $y_{2},y_{4}$ and $y_{5}$ can be rotated away if
\be
p_{L} &=& p_{y_{2}} +p_{E}~,~
p_{L_{S}} =p_{y_{4}}+p_{E}~,~p_{E_{S}}=
-p_{y_{5}}+p_{y_{2}} +p_{E}
\ee
are satisfied. So, only one free  phase is left,
which we assume to be $p_{L}$.
Then we cancel the phase of $h_{2}$
by $p_{L}$, which implies
\be
p_{L} &=& p_{h_{2}}.
\ee
No further phase rotation is possible, so that
$h_{3}$ and $h_{4}$ can be complex in general.
(Since for the mass matrices one can rotate 
the left-handed neutrinos and left-handed charged leptons
separately, one can eliminate one more
phase so that the neutrino mass matrix has only one 
independent phase.)
However, as we will see in the following discussions
that only the phase difference $
p_{h_3}-p_{h_4}$ enters into the  mass matrix
of the left-handed neutrinos and into the CP asymmetries
responsible for leptogenesis.

\subsection{Higgs sector}
Before we come to discuss  the 
double beta decay of the present model, we would like to
briefly summarize the feature of the Higgs sector.
The present model contains five  neutral  physical Higgs fields;
two scalars and three pseudo scalars.
Their couplings to the
fermions are basically fixed \cite{kubo1},
but the Yukawa sector does not satisfy the general
conditions \cite{glashow,paschos0} to suppress
the tree level FCNCs. The only way 
to suppress the tree level FCNCs in the model is to make
the Higgs particles sufficiently heavy $\gsim 10$ TeV, which mediate 
the tree level FCNCs. So, it is important to study
the Higgs potential.
The  $S_{3}$ invariant Higgs
potential $V_{H}$ has been studied in \cite{pakvasa1,kubo4}.
It has tuned out that  all the Higgs masses 
obtained from $V_{H}$ are proportional
to VEVs, so that
unless one discards the triviality constraint, they can be at most
of  the order of several hundreds GeV.
These values are too small to 
suppress FCNCs \cite{pakvasa1,yamanaka}.

One of the way out of the problem is to break
the $S_{3}$ symmetry softly;
as soft as possible to preserve the prediction 
from $S_{3}\times Z_2$ in the Yukawa sector.
It has been observed that if the Higgs potential  $V_{H}$
respects   $S_{3}$ as well as $Z_2$ invariance
($Z_2$ is defined in (\ref{z2})),
it has an additional abelian discrete symmetry
$S_2'$
\be
H_{1} &\leftrightarrow & H_{2},
\label{s2p}
\ee
 which is not a subgroup of the original $S_3$.
 Therefore, we assume that  the soft breaking mass term
 $\hat{V}_{SB}$ also respects this
 discrete symmetry $S_2'$, while breaking
 $S_3\times Z_2$ softly. The most general form is
 \be
\hat{V}_{SB} &=&
-\mu_{SB1}^2(H_1^{\dagger}H_2+h.c)-
\mu_{SB2}^2[~H_S^{\dagger}(H_{1}+H_2)+h.c.~].
\label{vh1}
\ee
It has been shown in \cite{kubo4} that
for the $S_3\times Z_2\times S_2'$
invariant Higgs potential with (\ref{vh1}), only
$S_2'$ invariant VEVs  (under the assumption
that $<H_{S}> \neq$ and $\mu_{SB1,2}$ are real)
\be
<H_{S}> &\neq &0, ~<H_{1}>=<H_{2}>\neq 0.
\label{s2vev}
\ee
can satisfy the condition that
 all the physical Higgs bosons, except one neutral physical Higgs boson, can
become heavy $\gsim 10$ TeV without having a problem
with triviality.
We would like to emphasize 
that the $S_{2}'$ invariant
 VEVs (\ref{s2vev}) are the most economic VEVs
in the sense that
the freedom of VEVs 
can be completely absorbed into the Yukawa couplings
so that we can derive the most general form for the fermion mass matrices
\be
{\bf M} = \left( \begin{array}{ccc}
m_1+m_{2} & m_{2} & m_{5} 
\\  m_{2} & m_1-m_{2} &m_{5}
  \\ m_{4} & m_{4}&  m_3
\end{array}\right)
\label{general-m}
\ee
without referring to the details of the Higgs potential.

To diagonalize  the Higgs fields,
we redefine the Higgs fields as
\be
H_{\pm} &=&\frac{1}{\sqrt{2}}(H_1\pm H_2),\nn\\
H_{L} &=&\cos\gamma H_{S}
+\sin\gamma H_{+}~,~
H_{H} = -\sin\gamma H_{S}
+\cos\gamma H_{+}
\ee
and 
\be
H_{-} &=& 
 \left( \begin{array}{c}
h_{-}\\
\frac{1}{\sqrt{2}} (h_{-}^{0}
  +i \chi_{-}) \\
\end{array}\right),~
  H_{L} = \left( \begin{array}{c}
h_{L} \\
 \frac{1}{\sqrt{2}} ( v+h_{L}^{0} +i \chi_{L}) \\
\end{array}\right),
 \label{hL}\\
H_H &=&
 \left( \begin{array}{c}
h_H \\
 \frac{1}{\sqrt{2}} (h_H^0+i \chi_H) \\
\end{array}\right)
   \label{hH},
\ee
where
\be
v_{+} &=& <h_{+}^{0}>~,~v_{S} = <h_{S}^{0}>~,~
v=(v_+^2+v_S^2)^{1/2}=246~\mbox{GeV},\nn\\
\sin \gamma &=&v_{+}/v~,~\cos\gamma =v_{S}/v.
\label{vevs}
\ee
As we see from (\ref{hL}), only   $H_{L}$ has VEV, and therefore, one 
can identify  $H_{L}$ as the SM Higgs doublet.
In fact, $h_{L}$ and $\chi_{L}$ are the would-be
Goldstone bosons. However, the neutral Higgs
$h_{L}^{0}$ is not a mass eigenstate; it mixes with
$h_{H}^{0}$. The mixing is of $O(v^{2}/\mu_{SB}^{2})$
which is at most $\sim 10^{-3}$. It is possible to
kill this mixing by fine tuning of the
couplings  in the Higgs potential.
In the following we assume this.
Under this assumption, all the Higgs fields defined 
in (\ref{hL}) and (\ref{hH}) are mass eigenstates.
In Table 1, their masses are given under the assumption
that $\mu_{SB1}^{2},\mu_{SB2}^{2} >> v^{2}$.

\vspace{0.3cm}
\begin{table}[htb]
\begin{center}
\begin{tabular}{|c|c|}
\hline
Higgs &  mass
\\ \hline
$h_{-}$ & $m_{h_{-}}^{2}\simeq 2\mu_{SB1}^{2}+
\sqrt{2}\mu_{SB2}^{2}\cot \gamma$
\\ \hline
$h_{-}^{0}$ & $m_{h_{-}^{0}}^{2}\simeq m_{h_{-}}^{2}$
\\ \hline
$\chi_{-}$ & $m_{\chi_{-}}^{2}\simeq m_{h_{-}}^{2}$
\\ \hline
$h_{L}$ & Would-be Goldstone
\\ \hline
$h_{L}^{0}$ & $m_{h_{L}}^{2}=O(v^{2})$
\\ \hline
$\chi_{L}$ & Would-be Goldstone
\\ \hline
$h_{H}$ & $m_{h_{H}}^{2}\simeq 2\sqrt{2}\mu_{SB2}^{2}/\sin 2\gamma$
\\ \hline
$h_{H}^{0}$ & $m_{h_{H}^{0}}^{2}\simeq m_{h_{H}}^{2}$
\\ \hline
$\chi_{H}$ & $m_{\chi_{H}}^{2}\simeq m_{h_{H}}^{2}$
\\ \hline
\end{tabular}
\caption{\label{table1}
Mass of the Higgs particles. $m_{h_{-,H}}$ should be
larger than $\sim 10$ TeV to
sufficiently suppress  the tree level FCNCs. }
\end{center}
\end{table}

\section{Neutrino mixing and neutrinoless double $\beta$ decay}
The fermion masses are generated from the $S_{2}'$
invariant VEVs (\ref{vevs}).
Because of the $Z_2$ symmetry (\ref{z2}), the mass matrix for the 
charged leptons becomes
\be
{\bf M}_{e} = \left( \begin{array}{ccc}
m_{2} & m_{2} & m_{5} 
\\  m_{2} & -m_{2} &m_{5}
  \\ m_{4} & m_{4}&  0
\end{array}\right),
\label{mlepton}
\ee
where
\be
m_{2} &=& v y_{2}  \sin\gamma
/\sqrt{2}, m_{4}=v  y_{4} \sin\gamma /\sqrt{2},
m_{5}=v y_{5} \sin\gamma  /\sqrt{2}.
\ee
As discussed previously,  the phase of all the
nonvanishing Yukawa couplings $y_{2},y_{4}$ and
$y_{5}$ can be rotated away. So, all the mass parameters
appearing in (\ref{mlepton}) are real.
Diagonalization of the mass matrices is straightforward.
The mass eigen values are
approximately given by
\be
m_e^2 &=& \frac{(m_{4} 
m_{5})^2}{(m_{2})^2+(m_{5})^2}
+O((m_{4})^4),\\
~m_\mu^2 &=& 2 (m_{2})^2+
(m_{4})^2+O((m_{4})^4),\\
m_{\tau}^2 &=&  2[~(m_{2})^2+(m_{5})^2~]+
\frac{(m_{4} m_2)^2}{(m_2)^2
+(m_{5})^2}+O((m_{4})^4).
\ee
Concrete values are given as
$m_4/m_5\simeq 0.00041$ and $ m_2/m_5 \simeq 0.0596$ and
$m_{5}\simeq  1254$ MeV
to obtain 
$m_e=0.51$ MeV, $m_\mu =105.7$ MeV and $m_\tau
=1777$ MeV.
The diagonalizing unitary matrices
(i.e., $U_{eL}^{T} {\bf M}_{e} U_{eR}$)
assume a simple form in the $m_{e} \to 0$ limit,
which is equivalent to the $m_{4} \to 0$ limit.
$U_{eL}$ in this limit is
\be
U_{eL}^{0} &= &\left( \begin{array}{ccc}
0 & -1/\sqrt{2} &  1/\sqrt{2}
\\  0 & 1/\sqrt{2}  & 1/\sqrt{2}
  \\ 1 & 0 &  0
\end{array}\right).
\label{UeL2}
\ee
We shall consider this limit later on.

Similarly, the Dirac neutrino mass matrix  is given by
\be
{\bf M}_{D} = \left( \begin{array}{ccc}
m_{D2} & m_{D2} & 0
\\  m_{D2} & -m_{D2} &0
  \\ m_{D4} & m_{D4}&   m_{D3}
\end{array}\right).
\label{mD}
\ee
where
\be
m_{D2} &=& v h_{2} \sin\gamma
 /\sqrt{2}, m_{D4}=v h_{4} e^{i p_{h_{4}}}\sin\gamma  /\sqrt{2},\nn\\
m_{D3} &= &v\cos\gamma h_{3}e^{ip_{h_{3}}}.
\label{mD2}
\ee
For the mass matrices one can rotate the left-handed charged 
leptons and the left-handed neutrinos separately.
So, we rotate $\nu_{SL}$ to absorb the phase of
$m_{D_4}$, and 
we rewrite ${\bf M}_{D}$ as
\be
{\bf M}_{D} = \left( \begin{array}{ccc}
m_{D2} & m_{D2} & 0
\\  m_{D2} & -m_{D2} &0
  \\ m_{D4} & m_{D4}&   m_{D3}\exp i \varphi_{3}
  \end{array}\right),
\label{mD3}
\ee
where
\be
\varphi_{3} &=&p_{h_{4}}-p_{h_{3}}~
(-\pi/2 \leq \varphi_{3} \leq \pi/2),
\label{phi3}
\ee
and the Dirac mass parameters 
$m_{D}$'s in (\ref{mD3}) are all real numbers.

The  Majorana masses for $\nu_L$ can be obtained from 
the see-saw mechanism,
and the corresponding mass matrix is given by $
{\bf M_{\nu}} = {\bf M_{D}}\tilde{{\bf M}}^{-1} 
({\bf M_{D}})^T$,
where $\tilde{{\bf M}}=\mbox{diag}(M_1,M_1,M_S)$.
We have assumed that the phases of the right-handed
neutrinos are used to rotate away the phase of 
$M_{1}$ and $M_{S}$. So, we may assume that
they are real positive numbers.
To express ${\bf M_{\nu}}$ in
a simple form we rescale the
Dirac neutrino masses according to
\be
m_{D2} & \to & \rho_{2}=m_{D2} /\sqrt{M_1}~,~
m_{D4} \to \rho_4 = m_{D4} /\sqrt{M_1},\nn\\
m_{D3} &\to &  \rho_3 =m_{D3} /\sqrt{M_S}.
\label{rescale}
\ee
Thanks to the $Z_2$ symmetry (\ref{z2}), the  mass matrix ${\bf M_{\nu}}$ 
 takes a simple form
\be
{\bf M}_{\nu} & = &{\bf M_{D}}\tilde{{\bf M}}^{-1} 
({\bf M_{D}})^T\nn\\
&=& \left( \begin{array}{ccc}
2 (\rho_{2})^2 & 0 & 
2 \rho_2 \rho_{4}
\\ 0 & 2 (\rho_{2})^2 & 0
  \\ 2 \rho_2 \rho_{4} & 0  &  
2 (\rho_{4})^2 +
(\rho_3)^2\exp i 2 \varphi_{3}
\end{array}\right).
\label{m-nu}
\ee
The $\rho$'s in (\ref{m-nu}) are real numbers.
One can convince oneself that ${\bf M}_{\nu}$ can be diagonalized as \cite{kubo2}
\be
U^T_\nu {\bf M}_{\nu} U_\nu &=& \left( \begin{array}{ccc}
m_{\nu_1}e^{i\phi_1-i\phi_\nu} & 0 & 0\\
0 & m_{\nu_2}e^{i\phi_2+i\phi_\nu} &0 \\
0 & 0 & m_{\nu_3}
\end{array}\right),
\ee
where
\be
U_{\nu}&= &\left( \begin{array}{ccc}
-s_{12} & c_{12}e^{i \phi_\nu}
&  0
\\ 0 & 0 &1
\\    c_{12}e^{-i \phi_\nu}  & s_{12}& 0
 \end{array}\right),\\
\label{unumax3}
m_{\nu_3} \sin \phi_\nu &=& m_{\nu_2} \sin \phi_2
=m_{\nu_1} \sin \phi_1~,~2 \varphi_{3}=\phi_{1}+\phi_{2}\\
\label{sinp}
m_{\nu_3} &=& 2 \rho_{2}^{2},~
\frac{m_{\nu_1}m_{\nu_2}}{m_{\nu_3}}= \rho_{3}^{2},
\label{rho23}\\
\tan \phi_\nu &=&
\frac{\rho_3^2 \sin2\varphi_3}{2(\rho_2^2+\rho_4^2)
+\rho_3^2\cos2\varphi_3},
\label{tanvarphi}
\ee
and $c_{12}=\cos\theta_{12}$ and $s_{12}=\sin\theta_{12}$.
We also find that
\be
\tan^2\theta_{12} &=&
\frac{(m_{\nu_2}^2-m_{\nu_3}^2 \sin^2\phi_\nu)^{1/2}
-m_{\nu_3}|\cos\phi_\nu|}{(m_{\nu_1}^2
-m_{\nu_3}^2 \sin^2\phi_\nu)^{1/2}
+m_{\nu_3}|\cos\phi_\nu|},
\ee
from which we find
\be
\frac{m_{\nu_2}^2}{\Delta m_{23}^2} &=&
\frac{(1+2 t_{12}^2+t_{12}^4-r t_{12}^4)^2}
{4  t_{12}^2 (1+t_{12}^2)(1+t_{12}^2-r t_{12}^2)\cos^2 \phi_\nu}
-\tan^2 \phi_\nu
\label{mnu2}\\
&\simeq &
\frac{1}{\sin^2 2\theta_{12}\cos^2 \phi_\nu}
-\tan^2 \phi_\nu ~~\mbox{for}~~|r| << 1,
\label{mnu21}
\ee
where $t_{12}=\tan\theta_{12}, r=\Delta m_{21}^2/\Delta m_{23}^2$.
It can also be shown  that  only an inverted mass spectrum
\be
m_{\nu_3} & < & m_{\nu_1}, m_{\nu_2}
\label{spectrum}
\ee
is consistent with  the experimental constraint $ |\Delta m_{21}^2|
< |\Delta m_{23}^2|$  in the present model.
Note that Eq. (\ref{sinp}) is satisfied for
\be
2 \varphi_{3} &=& \phi_{1}+\phi_{2} \sim \pm\pi
\label{1plus2}
\ee
and NOT for $\phi_{1} \sim \phi_{2}$. That is,
if $2 \varphi_{3}  \sim +(-)\pi$, then
$\cos\phi_1 < (>) 0$ and $\cos\phi_2 > (<) 0$.
Now the product $U_{eL}^{\dag}P  U_\nu$ 
defines a neutrino mixing matrix $V_{\rm MNS} $,
where
\be
P &=&\left( \begin{array}{ccc}
1 & 0 &0\\
0 & 1 & 0\\
0 & 0 & e^{i p_{h_4}}
\end{array}\right).\nn
\ee
For our purpose it is sufficient to use
the approximate unitary matrix $U_{eL}^{0}$
 given in (\ref{UeL2}) which is obtained in the
 limit that the electron mass is zero.
 We denote the approximate
  neutrino mixing matrix by $V_{\rm MNS}^{0}$
  obtained from $U_{eL}^{0}$ and $ U_\nu$.
 The product $U_{eL}^{0\dag} P U_\nu$
 can be brought by an
appropriate  phase transformation  to
a popular form
\be
V_{\rm MNS}  &\simeq & V_{\rm MNS}^{0} =
\left( \begin{array}{ccc}
 c_{12} c_{13} & s_{12} c_{13} &   s_{13} e^{-i\delta}\\
  -s_{12} c_{23}  -c_{12} s_{23} s_{13}e^{i\delta} & 
   c_{12} c_{23}  -s_{12} s_{23} s_{13}e^{i \delta}  &   s_{23}c_{13} \\
 s_{12} s_{23}  -c_{12} c_{23} s_{13}e^{i \delta} & 
   -c_{12} s_{23}  -s_{12} c_{23} s_{13}e^{i \delta}  &   c_{23}c_{13}
\end{array}\right) \nn\\
& & \times
\left( \begin{array}{ccc}
1 & 0 & 0\\
0 & e^{i \alpha} &0 \\
0 & 0 & e^{i \beta}
\end{array}\right).
\ee
with
\be
s_{13} & =& 0~,~t_{23} = \frac{s_{23}}{c_{23}}=
1,\\
\sin 2 \alpha &=&\sin(\phi_1-\phi_2)\nn\\
& =&
\pm \frac{ m_{\nu_3}\sin\phi_\nu}{m_{\nu_1}m_{\nu_2}}
\left( \sqrt{m_{\nu_2}^2-m_{\nu_3}^2 \sin^2 \phi_\nu}+
\sqrt{m_{\nu_1}^2-m_{\nu_3}^2 \sin^2 \phi_\nu} \right)
\label{alpha}\\
&\simeq &  \pm 2 \sin\phi_\nu (m_{\nu_3}/m_{\nu_2})
\sqrt{1-(m_{\nu_3}/m_{\nu_2})^2 \sin^2\phi_\nu},\nn \\
\sin 2 \beta &=&\sin(\phi_1-\phi_\nu)\nn\\
& = &
\pm \frac{\sin\phi_\nu}{m_{\nu_1}}
\left(m_{\nu_3}
 \sqrt{1-\sin^2 \phi_\nu}+
 \sqrt{m_{\nu_1}^2-m_{\nu_3}^2 \sin^2 \phi_\nu}
\right )
\label{sinb}
\ee
for $2\varphi_2 \sim \pm \pi$,
where $\phi_1,\phi_2$ and $\phi_\nu$ are defined in (\ref{sinp})
\footnote{For a nonvanishing electron mass,
we have $s_{13} \simeq m_e/\sqrt{2}m_\mu\simeq 0.0034$
and $\delta=p_{h_4}-\phi_\nu$.
Unfortunately, this value of $s_13$ is too small
to be measured \cite{minakata}.}.

The effective Majorana mass $<m_{ee}>$ in neutrinoless
double $\beta$ decay is  given by
\be
<m_{ee}> &=& |~\sum_{i=1}^3m_{\nu_i} V_{ei}^2|
\simeq |m_{\nu_1}c_{12}^2+
m_{\nu_2}s_{12}^2 \exp i2\alpha~ |,
\label{mee}
\ee
where $\phi_{\nu}$ and $\alpha$ are given in
(\ref{sinp}) and (\ref{alpha}), respectively.
In Fig.~1 we  plot $<m_{ee}>$  as a function of $\sin \phi_\nu$
for $\sin^2\theta_{12}=0.3, \Delta m_{21}^2=6.9 \times10^{-5}$ eV$^2$ and
$\Delta m_{23}^2=1.4, 2.3, 3.0 \times 10^{-3}$ eV$^2$ \cite{maltoni3}.
As we can see from Fig.~1, the effective Majorana mass stays
at about its minimal value $<m_{ee}>_{\rm min}$ for a wide range of $\sin\phi_\nu$.
Since $<m_{ee}>_{\rm min}$ is 
approximately equal to
$\sqrt{\Delta m_{23}^2}/\sin2\theta_{12}
=   ( 0.034  - 0.069 ) ~~\mbox{eV}$,
it is consistent with recent experiments \cite{klapdor1,wmap}
and is within an accessible range of future experiments \cite{klapdor2}.

\begin{center}
\begin{figure}[htb]
\includegraphics*[width=0.6\textwidth]{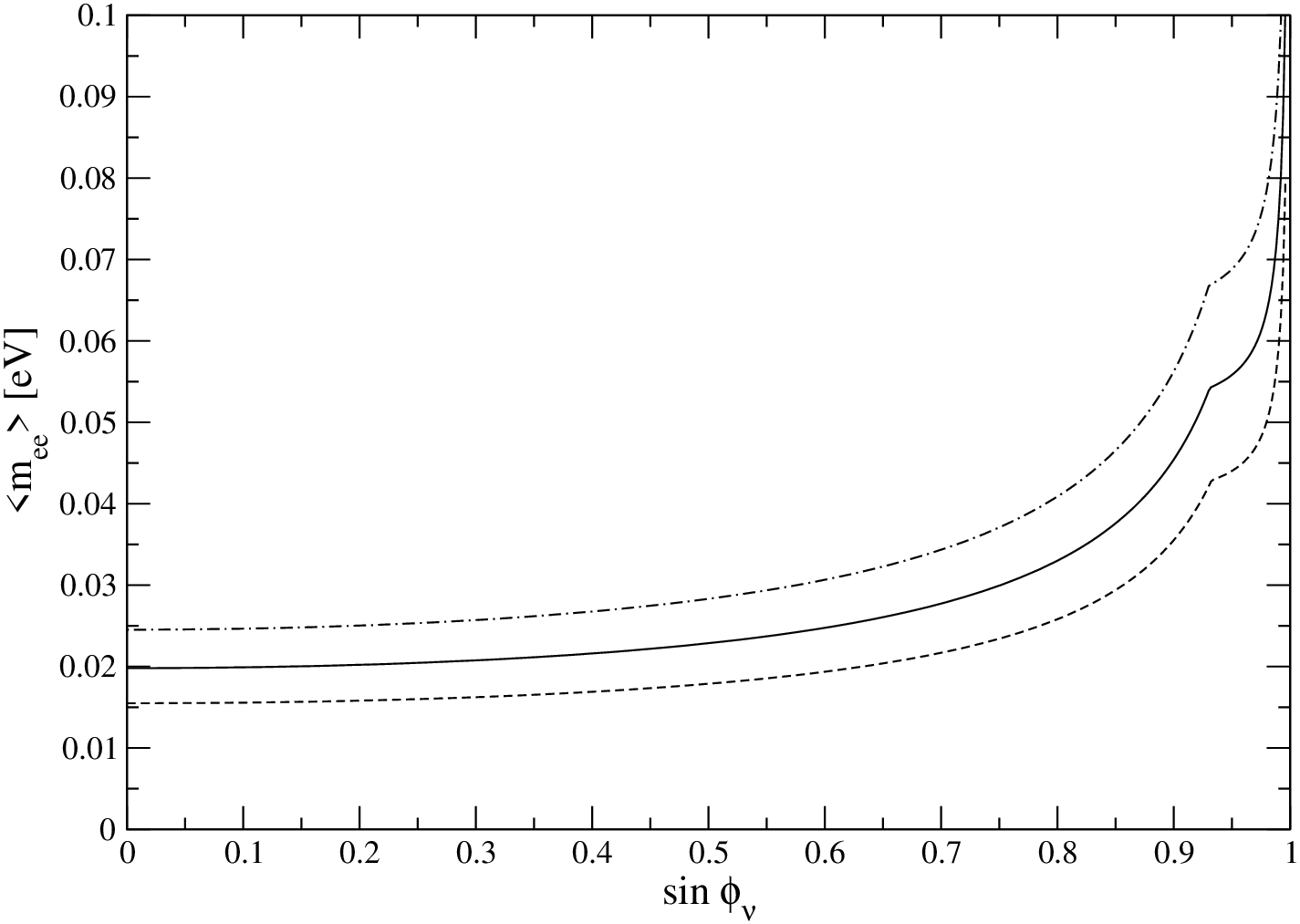}
\caption{\label{fig1}
The effective Majorana mass $<m_{ee}>$ as a function of
$\sin \phi_\nu$ with 
$\sin^2\theta_{12}=0.3$ and 
 $\Delta m_{21}^2=6.9 \times10^{-5}$ eV$^2$.
 The dashed, solid and dot-dashed lines stand for 
$\Delta m_{23}^2=1.4, 2.3$ and $ 3.0 \times 10^{-3}$ eV$^2$,
respectively. The $\Delta m_{21}^2$ dependence is very small.}
\end{figure}
\end{center}

Noticing that (\ref{sinp}), (\ref{1plus2})
and (\ref{alpha}), one obtains
\footnote{As we will see in the next section,
the ratio of baryons to photons $\eta_B$ is proportional to
$-\sin 2\varphi_3$.} (for $2\varphi_3 \sim  -\pi$)
\be
\sin 2 \varphi_{3} &=&
-\frac{m_{\nu_{3}}}{m_{\nu_{1}}} \sin\phi_\nu
\left[ 1-(\frac{m_{\nu_{3}}}{m_{\nu_{2}}}\sin\phi_\nu)^2\right]^{1/2}+
\frac{m_{\nu_{3}}}{m_{\nu_{2}}} \sin\phi_\nu
\left[ 1-(\frac{m_{\nu_{3}}}{m_{\nu_{1}}}\sin\phi_\nu)^2\right]^{1/2}\nn\\
&\simeq&
-\frac{m_{\nu_{3}}}{2m_{\nu_{2}}^{3}}
\frac{\Delta m_{21}^{2}\sin\phi_{\nu}}{
(1-(m_{\nu_{3}}/m_{\nu_{2}})^{2}\sin^{2}\phi_{\nu})^{1/2}},
\label{varphi3}
\ee
where $\Delta m_{21}^{2}/m_{\nu_{2}}^{2} <<1$ is assumed.
As one sees from (\ref{mnu2}), (\ref{spectrum}), (\ref{1plus2}), 
(\ref{alpha}) and (\ref{sinb}) that
once $\theta_{12}, \Delta m_{21}^{2}$
and $\Delta m_{23}^{2}$ are given, 
the only free parameter is $\phi_{\nu}$.
Numerically one finds
\be
\sin 2 \varphi_{3} &\simeq&
-(0.0034-0.013)\sin\phi_{\nu},
\label{varphi31}
\ee
where we have used:
$\sin^2\theta_{12}=0.3,
1.4~\mbox{eV}^2 \lsim \Delta m_{21}^2\times10^{5}
\lsim 3.0~\mbox{eV}^2 $ and\\
$6.1~\mbox{eV}^2 \lsim \Delta m_{23}^2\times 10^{3}
\lsim 8.4~\mbox{eV}^2$ \cite{maltoni3}.
Therefore, CP asymmetry being proportional
to $\sin 2 \varphi_3$ is very small in the present model,
even if the CP phase appearing in 
neutrinoless double $\beta$ decays is large \footnote{
Models in which the CP phases in the neutrino mixing
matrix are closely related to those 
for leptogenesis have been considered,
for instance, in \cite{frampton}-\cite{grimus5}.}.
\begin{center}
\begin{figure}[htb]
\includegraphics*[width=0.6\textwidth]{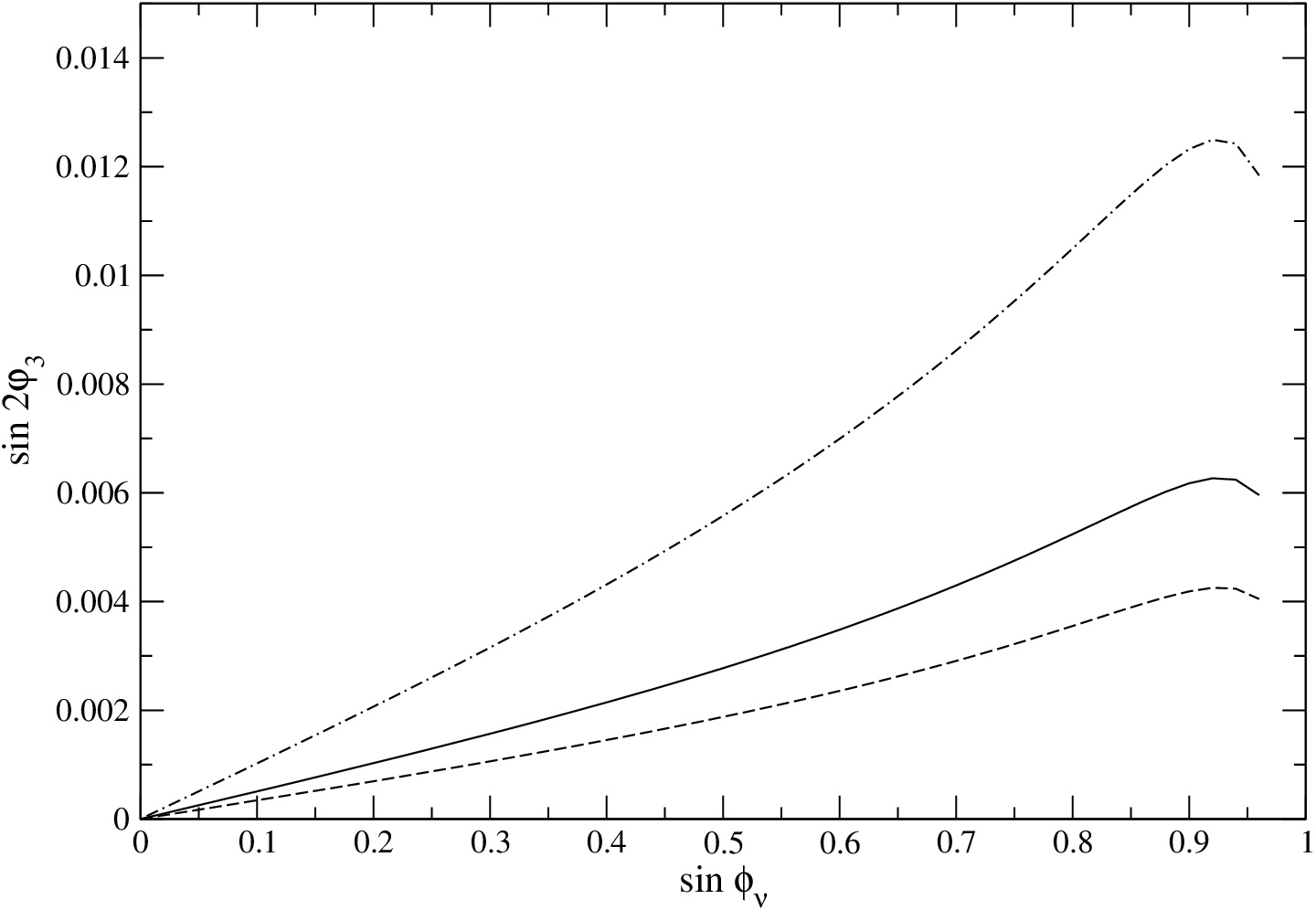}
\caption{\label{fig2}
$\sin 2 \varphi_{3}$ versus $\sin\phi_{\nu}$ 
with 
$\sin^2\theta_{12}=0.3$ in the case of $2\varphi_3 \sim +\pi$.
The case of $2\varphi_3 \sim -\pi$ is the same except for the sign
of $\sin 2\varphi_3$.
The dot-dashed, solid and dashed lines stand for 
$(\Delta m_{23}^2\times 10^{3}, \Delta m_{21}^2\times10^{5})
=(3.0, 6.1), (2.3, 6.9)$ and  $(1.4, 8.4)$ eV$^2$,
respectively.
}
\end{figure}
\end{center}

\section{Leptogenesis}
\subsection{CP phase}
Before we start to compute CP asymmetries, we first would like to
consider the mixing of the charged leptons in the 
limit that the mass of the electron vanishes. That is, 
we approximate the unitary matrix which
defines the mass eigenstates of the charged leptons by
$U_{eL}^0$ (\ref{UeL2}). 
We next rewrite the Yukawa interactions  (\ref{Ly}) 
in terms of the mass eigenstates 
for the Higgses (\ref{hL})-(\ref{hH})  and the charged leptons. 
The relevant part for  leptogenesis
becomes
\be
{\cal L}_{h} &=&
-h_{IJ}^{-}\overline{ \hat{L}}_I\epsilon  H_{-}^{*} \nu_{JR}
-h_{IJ}^{H}\overline{ \hat{L}}_I\epsilon  H_{H}^{*} \nu_{JR}
-h_{IJ}^{L}\overline{ \hat{L}}_I\epsilon  H_{L}^{*} \nu_{JR}
\label{Lh}
\ee
with
\be
\hat{L} &=& U_{eL}^0 L=  \left(  \left( \begin{array}{c}
\nu_{eL}\\
e_{L} \\
\end{array}\right)~,~
 \left( \begin{array}{c}
\nu_{\mu L}\\
\mu_{L} \\
\end{array}\right)~,~
\left( \begin{array}{c}
\nu_{\tau L}\\
\tau _{L} \\
\end{array}\right) \right),
\ee
where
\be
h_{IJ}^{H} &=&
\left(\begin{array}{ccc}
\cos\gamma h_{4}e^{i p_{h_{4}}}/\sqrt{2} & 
\cos\gamma h_{4}e^{i p_{h_{4}}}/\sqrt{2} &
-\sin\gamma h_{3} e^{i p_{h_{3}}}\\
0 & -\cos\gamma h_{2}&0 \\
\cos\gamma h_{2} & 0 & 0
\end{array}\right),
\label{yukH}\\
h_{IJ}^{L}& =&
\left(\begin{array}{ccc}
\sin\gamma h_{4}e^{i p_{h_{4}}}/\sqrt{2} & 
\sin\gamma h_{4}e^{i p_{h_{4}}}/\sqrt{2} &
\cos\gamma h_{3} e^{i p_{h_{3}}}\\
0 & -\sin\gamma h_{2}&0 \\
\sin\gamma h_{2} & 0 & 0
\end{array}\right),
\label{yukL}\\
h_{IJ}^{-} &=&
\left(\begin{array}{ccc}
h_{4}e^{i p_{h_{4}}}/\sqrt{2} & -h_{4}e^{i p_{h_{4}}}/\sqrt{2} &0 \\
h_{2} & 0 &0 \\
0 & h_{2} & 0
\end{array}\right),
\label{yukm}
\ee
where $\gamma$ is defined in (\ref{vevs}).

There are two types of diagrams that contribute to
CP asymmetries, vertex diagrams \cite{fukugita1,luty} 
and self-energy diagrams \cite{flanz1,flanz2}.
Nonvanishing CP asymmetries are proportional to
the imaginary part of 
$(h_{KI} h_{KJ}^{*} )^{2}$.
In the present case, there are three Higgs fields, and one finds that
if one neglects the mass difference of the Higgs bosons,
the vertex correction is proportional to the imaginary part of
\be
\sum_{J,K,M}~\sum_{A,B=H,L,-}~
(h_{JI}^A h_{JK}^{B*})(h_{MI}^B h_{MK}^{A*}),
\ee
while the self-energy correction is 
proportional to the imaginary part of
\be
\sum_{J,K,M}~\sum_{A,B=H,L,-}~
(h_{JI}^A h_{JK}^{A*})(h_{MI}^B h_{MK}^{B*}).
\ee
Since $h_{I1}$ and $h_{I2}$ have the same phase,
only the case $I=1,2 , K=3~\mbox{or}~I=3,K=1,2$ yields
nonvanishing CP asymmetries.
Therefore,  the matrix $h^{-}_{IJ}$ cannot contribute
to CP asymmetries. Moreover, since  $h_{i3}^{H,L}=0~(i=2,3)$,
both the vertex and self-energy contributions become
proportional to the imaginary part of 
\be
\sum_{A,B=H,L}
(h_{1i}^A h_{13}^{A*})(h_{1i}^B h_{13}^{B*}), i=1,2.
\ee
However, from 
 (\ref{yukH}) and (\ref{yukL}) we obtain
 \be
 h_{1i}^H h_{13}^{H*}+h_{iI}^Lh_{13}^{L*} &=&0, 
 \label{hh2}
\ee
 implying that CP asymmetries vanish,
if the mass differences of the Higgs bosons are neglected.
That is,  CP asymmetries, generated in one-loop
with $H_{L}$ and  $H_{H}$ exchanges, cancel
with each other.
In the presence of the 
soft $S_3\times Z_2$ breaking mass terms (\ref{vh1}), the  mass of $H_{H}$ 
can be considerably  different from that of $H_{L}$, as we can see from Table 1.
Consequently, 
there will be nonvanishing CP asymmetries
in a realistic case, in which $H_-$ and $H_H$ are made heavy
by the soft $S_3\times Z_2$ breaking terms to suppress the tree level FCNCs.

After so many discussions about the $S_3$
limit, we find that
\be
& & Im~[h_{1I }^{H,L}h_{1J}^{H,L *}]^{2}\sim
(h_{3}h_{4})^{2}\sin2(p_{h_{3}}-p_{h_{4}})(h_{3}h_{4})^{2}=
-(h_{3}h_{4})^{2}
\sin 2\varphi_{3}\label{hh}\\
& &\mbox{with}~~
I=1,2 , J=3~\mbox{or}~I=3,J=1,2,\nn
\ee
where $\varphi_{3}$ is given in (\ref{phi3}),
which is the only phase left over in the 
neutrino mass matrix. Therefore, the CP phase appearing
in the neutrino mixing is the same as that 
for CP asymmetries.

\subsection{CP asymmetries and Baryon number asymmetry}
We first calculate the total decay width of the right-handed neutrinos.
The relevant parts of the Yukawa interactions for this
purpose are given in (\ref{Lh})-(\ref{yukm}).
For the $S_{3}$ singlet right-handed neutrino $\nu_{SR}$ one finds
\be
\Gamma_{ST} &=&\Gamma_{S}[l+H_{L}]+
\Gamma_{S}[l+H_{H}]+\Gamma_{S}[l^{c}+H_{L}^{c}]+
\Gamma_{S}[l^{c}+H_{H}^{c}]\nn\\
&=& \frac{1}{8\pi}h_{3}^{2}M_{S}\left[
\cos^{2}\gamma+\sin^{2}\gamma
[~(1-\frac{m_{h_{H}}^{2}}{M_{S}^{2}})^2\theta(M_{S}-m_{h_{H}} )~]
\right]
\label{gammast}\\
& \to & \frac{1}{8\pi}h_{3}^{2}M_{S}~~\mbox{as}~~m_{h_{H}}/M_{S} \to 0, \nn\\
& \to &  \frac{1}{8\pi}h_{3}^{2}M_{S}
\cos^{2}\gamma=\frac{1}{8\pi}
\left(\frac{m_{\nu_1} m_{\nu_2} }{m_{\nu_3} }\right)
\left(\frac{M_S}{v^2}\right)^2~~
\mbox{as}~~m_{h_{H}}/M_{S}\to 1,\nn
\ee
where the first term result from the decay
into the SM Higgs  $H_{L}$, and
the second term comes from the decay
into  $H_{H}$ with mass
$m_{h_{H}}$. 
The last equality follows from (\ref{mD2}), (\ref{mD3}),
(\ref{rescale}) and  (\ref{rho23}).
Similarly, one finds the total decay width
of $\nu_{1R}$ and $\nu_{2R}$
\be
\Gamma_{1T} &=&\Gamma_{2T} \nn\\
&=&\frac{1}{8\pi}(\frac{1}{2}h_{4}^{2}+h_{2}^{2})M_{1}\left[
\sin^{2}\gamma
+\cos^{2}\gamma[~(1-\frac{m_{h_{H}}^{2}}{M_{1}^{2}})^2
\theta(M_{1}-m_{h_{H}} )~]\right.\nn\\
& &\left.+[~(1-\frac{m_{h_{-}}^{2}}{M_{1}^{2}})^2\theta(M_{1}-m_{h_{-}} )~]
\right]
\label{gamma1t}\\
& \to 
&\frac{1}{4\pi}(\frac{1}{2}h_{4}^{2}+h_{2}^{2})M_{1}~
~\mbox{as }~~m_{h_{H}}/M_{1} ,m_{h_{-}}/M_{1}\to 0\nn,\\
& \to &  \frac{1}{8\pi}(\frac{1}{2}h_{4}^{2}+h_{2}^{2})M_{1}
\sin^{2}\gamma =\frac{1}{8\pi}
(m_{\nu_3} +\rho_4^2)
\left(\frac{M_1}{v}\right)^2~~
\mbox{as}~~m_{h_{H}}/M_{1} ,m_{h_{-}}/M_{1}\to 1,\nn
\ee
where $M_{1}=M_{2}$ is assumed.

Because of (\ref{varphi31}), i.e.
$|\sin\varphi_3| \lsim 0.013$,  one needs an
enhancement to obtain a realistic value of baryon
asymmetry.  A nice way is the resonant 
enhancement \cite{flanz1,flanz2,pilaftsis1},
which we consider below \footnote{See, for instance, \cite{hambye1}-\cite{bando}
for recent models with resonant enhancement of leptogenesis.
See also \cite{pilaftsis2}.}.
Since in this case the self-energy contributions
to CP asymmetries dominate, we consider only them
and neglect the contributions coming from
the vertex diagrams.
In the $S_{3}$ symmetric limit,
$M_{1}$ is equal to $M_{2}$. This relation is modified
to $M_{1}=M_{2}+O(m_{\nu})$
because of spontaneous symmetry breaking of $S_3$.
So, there is  a natural degeneracy of 
 $\nu_{1R}$ and $\nu_{2R}$.
However,  there is no 
resonant enhancement
between $\nu_{1R}$ and $\nu_{2R}$,
because $\mbox{Im}h_{i1}^{H,L}(h_{i2}^{H,L})^*=0$
as we can see from (\ref{yukH}) and (\ref{yukL}).
Therefore, we may  neglect this small
correction, and we have to assume that
$M_{1} =M_{2}\simeq M_{S}$.
Introducing the notation
\be
\Delta M^{2}/M_{S}^{2}=1-\frac{M_{1}^{2}}{M_{S}^{2}}
=1-x
\sim 0,
\ee
we  find that
\be
\Delta \Gamma_{S} &=&
\Gamma_{S}[l+H_{L}]+
\Gamma_{S}[l+H_{H}]-\Gamma_{S}[l^{c}+H_{L}^{c}]-
\Gamma_{S}[l^{c}+H_{H}^{c}]
\nn\\
&=& \frac{1}{64\pi^2}(h_{4}h_3)^{2}(\sin^2\gamma \cos^2\gamma)
\sin 2\varphi_{3}
[ 1-(1-y_{H})^{2}\theta(1-y_{H})]^{2}
\frac{M_1}{1-x},
\label{deltagamma}
\ee
and $\Delta \Gamma_{1}\simeq \Delta \Gamma_{S}/2$
for $M_S\simeq M_1$.
We have assumed that
\be
\Gamma_{ST,1T} /M_{S} << |\Delta M^{2}/M_{S}^{2}|
\ee
to use the approximate formula (\ref{deltagamma}).
From these calculations we   obtain the CP asymmetries
\be
\epsilon_{S} &=&
\frac{\Delta \Gamma_{S}}{ \Gamma_{ST} }
\nn\\
&=& \frac{1}{8\pi}h_{4}^{2}
\sin 2\varphi_{3}\frac{
[ 1-(1-y_{H})^{2}\theta(1-y_{H})]^{2}}{[1/\sin^{2}\gamma
+ (1-y_{H})^{2}\theta(1-y_{H})/\cos^{2}\gamma]}~
\frac{\sqrt{x}}{1-x},
\label{cps}\\
\epsilon_{1} &=&\epsilon_{2}
\simeq  \frac{1}{2}\frac{\Gamma_{ST}}{\Gamma_{1T}}\epsilon_{S}
\label{cps1},
\ee
where 
\be
 y_{H}=\frac{m_{h_{H}}^{2}}{M_{S}^{2}}.
 \ee
From (\ref{cps}) various limits may be obtained:
\be
\epsilon_{S}& \to & 0~~
\mbox{as}~~y_{H} \to 0
\label{eslimit} \\
& \to &  \frac{1}{8\pi}h_{4}^{2}
\sin 2\varphi_{3} \sin^{2}\gamma \frac{\sqrt{x}}{1-x}
=
 \frac{1}{4\pi}\left(\frac{\rho_4^2 M_S}{v^2}\right)
\sin 2\varphi_{3} \frac{\sqrt{x}}{1-x}
~~
\mbox{as}~~y_{H}\to 1,
\ee
where we have used 
 (\ref{mD2}), (\ref{mD3})  and (\ref{rescale}).

To be definite  we assume $y_{H}, y_{-}
=m_{h_{-}}^{2}/M_{S}^{2} \sim 1$
in the following discussions.
Then only the SM Higgs $H_L$ contributes to 
CP asymmetries, and the phase $\phi_{\nu}$
(or $\varphi_3$),
$\sin^2\gamma$  (defined in (\ref{vevs}) )
and the effective mass
\be
M_{\rm eff} &=& \frac{M_{1}}{1-x} = 
\frac{M_{S}^{2}M_{1}}{M_{S}^{2}-M_{1}^{2}}
\simeq \frac{M_{S}^{3}}{M_{S}^{2}-M_{1}^{2}}
\label{meff}
\ee
 are the only independent parameters.
 The lepton and baryon asymmetries $Y_L=n_L/s$ and $Y_B=n_B/s$
 ($n_L, n_B$ are the lepton and baryon number density, and 
$s$ is the entropy density)
are given by \cite{khlebnikov}
\be
Y_{L} & \simeq & \kappa_{S}\epsilon_{S}/g^{*}+
2\kappa_{1}\epsilon_{1}/g^{*} ~\mbox{with}~g^{*}\simeq 120,
\label{yL}\\
Y_{B} &=&\frac{\omega}{\omega-1}Y_{L},
\label{yB}\\
\omega &=&
\frac{8 N_{F}+4 N_{H}}{22 N_{F}+13 N_{H}}\simeq 0.34
~\mbox{for}~~N_{F}=3, N_{H}=3,
\ee
where $k_{S(1)}$ is the dilution factor for the CP asymmetry $\epsilon_{S(1)}$,
and $g^*$ is the effective number of degrees of freedom
at the temperature $T=M_S \simeq M_1=M_2$.
We have taken into account all the degrees of freedom
in $g^*$
including three right-handed neutrinos and three Higgs doublets.
The dilution factors can be approximately written as
\cite{kolb,flanz3,pilaftsis1}
\be
\kappa_{S} &\simeq & \frac{0.3}{K_S
[\ln K_S]^{3/5} },~\kappa_{1} \simeq  \frac{0.3}{K_1
[\ln K_1]^{3/5} },
\label{ditution}
\ee
where
\be
K_{S} &=& \Gamma_{ST}/H_{ST}
=\frac{h_{3}^{2}}{16 \pi }\frac{M_{PL}}{1.66\sqrt{g^{*}} M_{S}}
=\frac{\rho_{3}^{2}}{8 \pi v^{2}}
\frac{M_{PL}}{1.66\sqrt{g^{*}}}\\
&\simeq & 4.4\times 10^{2}
\frac{(m_{\nu_{1}}/ 1~\mbox{eV}) (m_{\nu_{2}}/ 1~\mbox{eV})}
{(m_{\nu_{3}}/1~\mbox{eV})}
\label{ks}\\
K_{1} &=&\frac{\rho_{4}^{2}+m_{\nu_{3}}}{8 \pi v^{2}}
\frac{M_{PL}}{1.66\sqrt{g^{*}}}
 \simeq 4.4 \times 10^{2}(\rho_{4}^{2}+m_{\nu_{3}})/1~\mbox{eV},\label{k1}
\ee
where $\rho$'s are given in (\ref{rescale}), and (\ref{rho23}) is used.
[The approximate formula (\ref{ditution}) is applicable for
$10 \lsim K_{S,1} \lsim  10^6$.]
Note that using (\ref{rho23}) and (\ref{tanvarphi}),
we can express $\rho_{4}^{2}$ in terms of the neutrino masses, $\phi_\nu$
and $\varphi_3$, and  we find that the $\sin\gamma$ dependence
in $\kappa_S$ and $\kappa_1$ cancels so that the lepton asymmetry
and hence the baryon asymmetry $Y_B$  does not depend on
$\sin\gamma$.
\begin{center}
\begin{figure}[htb]
\includegraphics*[width=0.6\textwidth]{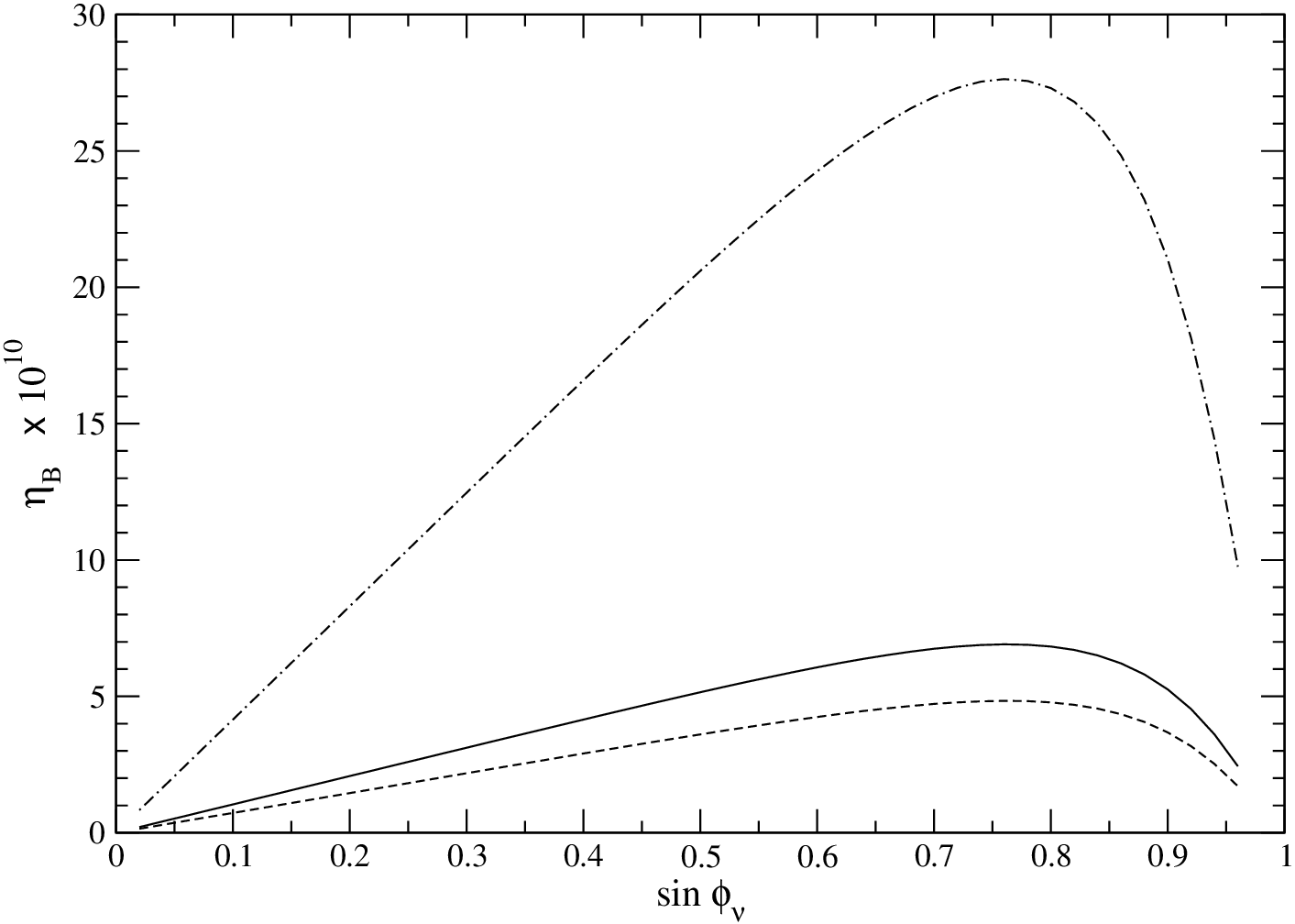}
\caption{\label{fig3}$\eta_B$ versus $\sin\phi_\nu$.
The dot-dashed, solid and doted lines correspond
to $M_{\rm eff}=M_1/(1-x)=4.0, 1.0, 0.7 \times 10^{13}$ GeV.
We have assumed that $m_{h_H}, m_{h_-} \simeq M_S$ and used
$\sin^2\theta_{12}=0.3, \Delta m^2_{23}
=2.3 \times 10^{-3}~\mbox{eV}^2,  
\Delta m^2_{21}=6.9 \times 10^{-5}~\mbox{eV}^2$.
The experimental value of $\eta_B \times 10^{10}$
is about  $6.5$ \cite{wmap}.}
\end{figure}
\end{center}
Finally, the ratio of the baryon number density to the photon density $\eta_{B}$
is given by
\be
\eta_{B} &\simeq &7.04 Y_{B}
\simeq -3.0 \times 10^{-2}
(\kappa_{S}\epsilon_{S}+
2\kappa_{1}\epsilon_{1}).
\label{etaB}
\ee
In Fig.~3
$\eta_{B}$ as a function of $\sin\phi_{\nu}$
is plotted for three different  values of  
$M_{\rm eff}=M_1/(1-x)=4.0, 1.0, 0.7 \times 10^{13}$ GeV.
As we see from Fig.~ 3,
 $\eta_{B}$ becomes maximal about
$\sin\phi_{\nu}\simeq 0.75$.
To obtain a realistic value of $n_{B}(\simeq 6.5 \times 10^{-10})$,
the effective mass $|M_{\rm eff}|=M_{1}/|1-x|$
should be of $0(10^{13})$ GeV, which means that $|1-x|$ has to be very small
 if $M_{1}$ is much smaller than $10^{13}$ GeV.
We have to fine tune $M_{S}$ so that 
$|1-x|=|1-M_{1}^{2}/M_{S}^{2}|\simeq 10^{-9}$
for $M_{S}=10$ TeV, for instance.

Let us at last discuss how fine this fine tuning is.
A fine tuning is unnatural, if radiative corrections 
are larger than the fine tuning.
One-loop radiative correction to the
the right-handed neutrino masses
may be estimated to be  $\delta M\sim M h^{2}/16 \pi^{2}$, where $h$ 
stands for the generic Yukawa couplings in (\ref{Ly}),
and $M$ for $M_1$ and $M_S$.
So, the fine tuning of $(1-x)$ will be natural if
\be
|1-x | &= &| 1-M_{1}^{2}/M_{S}^{2}|
>  \delta M/M\sim  h^{2}/16 \pi^{2}.
\label{natural}
\ee
This condition, however, is weaker than the condition,
$|M_{S}^{2}-M_{1}^{2} | >>
M_{S}\Gamma_{ST}$ and $M_1 \Gamma_{1T}$,  for the
approximate formula (\ref{cps})
for one-loop self-energy diagram to be
applicable \cite{pilaftsis1}.
The later condition  is equivalent to
\be
|1-x |
 & >>& \frac{1}{8 \pi}
 ( h_{2}^{2}\sin^2\gamma, h_{2}^{2}\sin^2\gamma, 
 h_{3}^{2}\cos^2\gamma),
 \label{cond3}
 \ee 
as we can see from (\ref{gammast}) and (\ref{gamma1t}).
Therefore, if this condition is satisfied, the naturalness
condition is automatically satisfied.
The value of $h$ can be estimated
from (\ref{mD2}), (\ref{mD3}), (\ref{rescale}) and (\ref{rho23}):
\be
 \frac{1}{8 \pi} h_{2}^{2}\sin^2\gamma
 &=&\frac{1}{8\pi}
 \left(\frac{m_{\nu_3}}{v}\right)\left( \frac{M_1}{v}\right)
 \lsim 10^{-13} \frac{M_1}{v},\\
 \frac{1}{8 \pi} h_{4}^{2}\sin^2\gamma &=&
\frac{1}{16\pi}\left( \frac{m_{\nu_1}m_{\nu_2}}{m_{\nu_3}v}
 \frac{\sin 2 \varphi _3}{\tan\phi_\nu}
 + \frac{m_{\nu_1}m_{\nu_2}}{m_{\nu_3}v}\cos 2\varphi_3
-\frac{m_{\nu_3}}{v}\right) \left(\frac{M_{1}}{v}\right)\nn\\
& \lsim & 10^{-13} \frac{M_1}{v},\\
 \frac{1}{8 \pi} h_{3}^{2}\cos^2\gamma &= &
 \frac{1}{8 \pi}
 \left(\frac{ m_{\nu_{1}}m_{\nu_{2}}/m_{\nu_{3}}}{v}\right)
 \left(\frac{M_{S}}{v}\right)
  \lsim 10^{-13} \frac{M_S}{v}.
 \ee
The last inequality is obtained as follows.
First we use the fact that in the present model
an inverted hierarchy is predicted, and the experimental bound
\cite{wmap} $m_{\nu} \lsim 0.2$ eV.
The ratio $|\sin 2\varphi_3/\tan\phi_{\nu}|$ is 
less than $0.013$ because of (\ref{varphi31}).
Further from (\ref{mnu21}) we  find
 \be
 m_{\nu_2}^2 \lsim \frac{\Delta m_{23}^2}{\sin^2 2 \theta_{12}},
\ee
which gives
\be
\frac{m_{\nu_2}^2}{m_{\nu_3}^2}
=\left( 1-\frac{\Delta m_{23}^2}{m_{\nu_2}^2}\right)^{-1}
\lsim \left( 1-\sin^2 2\theta_{12}\right)^{-1}
=\cos^{-2} 2\theta_{12} \lsim (4.6)^2.
\ee
Therefore,  the condition (\ref{cond3})  becomes
\be
|1-x| & >> &10^{-13}\frac{M_{1}}{v}.
\ee
In terms of the effective mass $M_{\rm eff}$ one finally finds
\be
|M_{\rm eff}| &=&\frac{M_{1}}{|1-x|}
<<10^{13} v \simeq  10^{15} ~\mbox{GeV}.
\label{cond4}
\ee
Therefore, 
the criterion on the validity of the approximate formula (\ref{cps}) 
does not depend on the mass of the right-handed
neutrinos. We recall that if (\ref{cond4}) is satisfied, 
the naturalness condition  (\ref{natural})
is automatically satisfied.
For $M_{S}=10$ TeV, for instance,
we obtain $|1-x |>> 4 \times 10^{-13}$
 which implies that the fine tuning of
 $|1-x |\simeq 10^{-9}$ to obtain $M_{\rm eff} \simeq 10^{13}$ GeV
 is not only unnatural, but also the use of the approximate formula
 (\ref{cps})  is justified.
 The main reason of the independence 
of the right-handed
neutrinos masses, $M_1$ and $M_S$, is the see-saw mechanism;
the smaller $M_1$ and $M_S$ are, the finer fine tuning
of $(1-x)$ is allowed because of (\ref{cond3}).

As we can see from (\ref{eslimit}),
the CP asymmetries in the present model
vanish if the Higgs masses are much smaller than
the right-handed
neutrinos masses. On one hand, the heavier the Higgs masses are,
the finer fine tuning is needed in the Higgs sector. 
The constraints coming from FCNCs, on the other hand,
require them to be larger than $O(10)$ TeV \cite{pakvasa1,yamanaka}.
Therefore, if we would like to explain the observed baryon asymmetry
from leptogenesis within the framework of the present
model, it is theoretically desirable to
have right-handed
neutrinos masses less than, say,  $O(100)$ TeV.

\section{conclusion}
In this paper we considered a minimal
 $S_3$ extension of the SM and investigated the possibility
 to explain the observed baryon asymmetry 
 in the universe through leptogenesis \cite{fukugita1}.
Below we would like to summarize our findings.\\

\vspace*{0.5cm}
\noindent
1. As in \cite{kubo1,kubo2}, we assumed an additional 
discrete symmetry (\ref{z2}) to increase the predictive
power in the leptonic sector. 
The leptonic sector of the Yukawa interactions 
contains two independent phases
$p_{h_4}$ and $p_{h_3}$.
We found that in the limit that the electron mass vanishes,
only the combination $2\varphi_3=
2(p_{h_3}-p_{h_4})$ enters into the neutrino mixing matrix $V_{\rm MNS}$
as well as into the CP asymmetries $\epsilon$'s
responsible for leptogenesis. (In this limit,
$V_{e3}$ vanishes.)
However, because of (\ref{varphi31}), we obtained
$\delta_{CP}= |\sin 2\varphi_3| \lsim 0.013$. 

\vspace*{0.5cm}
\noindent
2. It turned out that in the $S_3$ symmetric limit, the CP asymmetries vanish.
Therefore,  within the framework of the minimal extension,
one has to break $S_3$ explicitly.
To keep the predictivity in the Yukawa sector, we broke it softly.
We note that the same soft masses were introduced in \cite{kubo4}
to make the heavy Higgses heavy $\gsim O(10)$ TeV in order to suppress
sufficiently the tree level FCNCs.

\vspace*{0.5cm}
\noindent
3. Because of $|\sin 2\varphi_3| \lsim 0.013$,
an enhancement is needed. 
A nice way is the resonant enhancement \cite{flanz1,flanz2,pilaftsis1},
which requires all the right-handed neutrino masses
are degenerate. [$M_1=M_2$, 
up to  very small corrections, is ensured by $S_3$ symmetry
even if it is softly broken.]

\vspace*{0.5cm}
\noindent
4.  We also found  that the CP asymmetries vanish
if the right-handed neutrino masses are much larger than
the heavy Higgs masses.
Therefore, to obtain a realistic size for baryon asymmetry,
we have to assume that the right-handed neutrino masses
are bounded from above. 
The heavy Higgs masses dictate
the upper bound.

\vspace*{0.5cm}
\noindent
5. At last, we investigated the question of 
how fine the fine tuning needed for the degeneracy
of the neutrino masses is.
We found that if the criterion  (\ref{cond4}) on the validity of the approximate 
formula (\ref{cps})  is satisfied, 
the naturalness condition  (\ref{natural})
is automatically satisfied.
 It turned out that the criterion  (\ref{cond4})
 does not depend on the mass of the right-handed
neutrinos.

\vspace*{0.5cm}
As Fig.~3 presents, it is possible in the present model to explain 
the observed baryon
asymmetry in the universe by leptogenesis.
The basic parameters are:
the right-handed neutrino masses, the heavy Higgs masses and the CP phase.
Since the resonant enhancement of CP asymmetries
is assumed, and consequently, the degeneracy among
 the right-handed neutrino masses has to be very precise,
it will be very difficult to experimentally determine the right-handed neutrino masses
from a  precise measurement of baryon asymmetry alone,
even if the CP phase is precisely known.
Experimentally, this is not a nice feature, but 
the model predicts (if the observed baryon asymmetry 
should be explained by  leptogenesis)
that there will be three extremely degenerate right-handed neutrinos
whose masses are comparable with or  less than the heavy Higgs masses.
On one hand, the smaller the heavy Higgs masses are,
the more natural is the fine tuning in the Higgs sector.
The masses of the heavy Higgses, on the other hand,
are $\gsim  O(10)$ TeV to sufficiently suppress
the tree level FCNCs.
Therefore, we may theoretically 
expect right-handed neutrino masses of $O(10)$ TeV
in the present model.

\end{document}